\documentclass[a4paper]{article}

\usepackage{INTERSPEECH2020}
\usepackage{multirow}

\title{WeNet: Production Oriented Streaming and Non-streaming End-to-End Speech Recognition Toolkit}
\name{Zhuoyuan Yao$^{1,3}$,  Di Wu$^{2,3}$, Xiong Wang$^1$, Binbin Zhang$^{2,3}$,  Fan Yu$^1$, Chao Yang$^{2,3}$, Zhendong Peng$^{2,3}$, Xiaoyu Chen$^{2,3}$, Lei Xie$^{1*}$\thanks{$^*$ Lei Xie is corresponding author}, Xin Lei$^2$}
\address{
  $^1$Audio, Speech and Language Processing Group (ASLP@NPU),
School of Computer Science, Northwestern Polytechnical University, Xi’an, China \\
  $^2$Mobvoi Inc., Beijing, China, $^3$WeNet Open Source Community 
  }
\email{zhyyao@npu-aslp.org
       di.wu@mobvoi.com 
       fyu@npu-aslp.org 
       binbinzhang@mobvoi.com 
       fyu@npu-aslp.org 
       chaoyang@mobvoi.com 
       zhendong.peng@mobvoi.com 
       xiaoyu.chen@mobvoi.com 
       lxie@nwpu.edu.cn 
       mikelei@mobvoi.com}

\begin{document}

\maketitle
\begin{abstract}
In this paper, we propose an open source speech recognition toolkit called WeNet, in which a new two-pass approach named U2 is implemented to unify streaming and non-streaming end-to-end (E2E) speech recognition in a single model. 
The main motivation of WeNet is to close the gap between the research and deployment of E2E speech recognition models. WeNet provides an efficient way to ship automatic speech recognition (ASR) applications in real-world scenarios, which is the main difference and advantage to other open source E2E speech recognition toolkits. We develop a hybird connectionist temporal classification (CTC)/attention architecture with transformer or conformer as encoder and an attention decoder to rescore th CTC hypotheses.
To achieve streaming and non-streaming in a unified model, we use a dynamic chunk-based attention strategy which allows the self-attention to focus on the right context with random length. 
Our experiments on the AISHELL-1 dataset show that our model achieves 5.03\% relative character error rate (CER) reduction in non-streaming ASR compared to a standard non-streaming transformer. After model quantification, our model achieves reasonable RTF and latency at runtime.
The toolkit is publicly available at https://github.com/mobvoi/wenet.
\end{abstract}


\section{Introduction}

End-to-end (E2E) automatic speech recognition (ASR) models have gained more and more attention over the last few years, such as connectionist temporal classification (CTC)~\cite{graves2006connectionist,amodei2016deep}, recurrent neural network transducer (RNN-T),~\cite{graves2012sequence,graves2013speech,wang2021cascade,wang2019exploring} and attention based encoder-decoder (AED)~\cite{chorowski2014end,chan2015listen,chorowski2015attention,luo2021simplified,miao2020transformer}. Compared with the conventional hybrid ASR framework, the most attractive merit of E2E models is the extremely simplified training procedure. 

Recent work ~\cite{prabhavalkar2017comparison,sainath2019two,kim2017joint} also shows that E2E systems have surpassed conventional hybrid ASR systems in the standard of word error rate (WER). Considering the foregoing mentioned advantages of E2E models, deploying the emerging ASR framework into real-world productions becomes necessary. However, deploying an E2E system is not easy and there are a lot of practical problems to be solved.

First, \textbf{the streaming problem}. Streaming inference is essential for many scenarios that require the ASR system to respond quickly with low latency. However, it is difficult for some E2E models to run in a streaming manner, such as LAS~\cite{chan2015listen} and Transformer~\cite{vaswani2017attention}. Either great effort is required or obvious accuracy loss is introduced to make such model work in the streaming fashion~\cite{raffel2017online,chiu2017monotonic,inaguma2020enhancing}.

Second, \textbf{unifying streaming and non-streaming modes}. Streaming and non-streaming systems are usually developed separately. Unifying streaming and non-streaming in a single model can reduce the development effort, training cost as well as deployment cost, which is also preferred for production adoption~\cite{yu2020universal,tripathi2020transformer,hu2020deliberation,hu2021transformer}.

Third, \textbf{the production problem}, which is the most important problem we care about during the WeNet design. Great efforts are required to promote the E2E model into a real production application. So we have to carefully design the inference workflow in terms of model architecture, applications and runtime platforms. Due to the working manner of autoregressive beam search decoding, the workflow of most E2E model architecture is extremely complicated. Also, the cost of computation and memory should be considered during the model deployment on edge devices.
As for runtime platforms, although there are various platforms can be used to do neural network inference, such as ONNX (Open Neural Network Exchange), LibTorch in Pytorch, TensorRT\cite{vanholder2016efficient}, OpenVINO, MNN\cite{alibaba2020mnn} and NCNN, it still requires both speech processing and advanced deep learning optimization knowledge to select the best one for specific applications.

In this work, we present WeNet to address the above problems. ``We" in WeNet is inspired by ``WeChat"\footnote{WeChat is the most popular instant messaging platform on mobile devices in Chinese society.}, which means connection and share, and ``Net" is from Espnet~\cite{watanabe2018espnet} since we have referred to a lot of excellent designs in Espnet. Espnet is the most popular open source platform for end-to-end speech research. It mainly focuses on end-to-end ASR, and adopts widely-used dynamic neural network toolkits Chainer and PyTorch as the main deep learning engine.
By contrast, the main motivation of WeNet is to close the gap between research and production of E2E speech recognition models. 
On the production oriented principles, WeNet adopts the following implementations.
First, we propose a new two-pass framework namely U2 to solve the unified streaming and non-streaming problem.
Second, from model training to deployment, WeNet only depends on PyTorch and it's ecosystem. The key advantages of WeNet are as follows.

\textbf{Production first and production ready}: The Python code of WeNet meets the requirements of TorchScript, so the model trained by WeNet can be directly exported by Torch Just In Time (JIT) and LibTorch is used for inference. There is no gap between research and production models. 

\textbf{Unified solution for streaming and non-streaming ASR}: WeNet adopts the U2 framework to achieve an accurate, fast and unified E2E model, which is favorable for industry adoption.

\textbf{Portable runtime}: Several runtimes are provided to show how to host WeNet trained models on different platforms, including server (x86) and embedded (ARM in Android platforms).

\textbf{Light weight}: WeNet is designed specifically for E2E speech recognition with clean and simple code and all based on PyTorch and its ecosystem. So it has no dependencies on Kaldi\cite{povey2011kaldi}, which simplifies installation and usage.

Our experiments has shown that, WeNet is a easy-learned speech recognition toolkit with an end-to-end solution from research to production. In this paper, we will describe the model architecture, system design and runtime benchmark including real-time factor (RTF) and latency.

\section{WeNet}

\subsection{Model Architecture}

As we aim to address the streaming, unifying and production problem, the solution should be simple and easy to build and convenient to be applied at runtime, while keeping good performance.

\begin{figure}[h]
  \vspace{-10pt}
  \centering
  \includegraphics[scale=0.8]{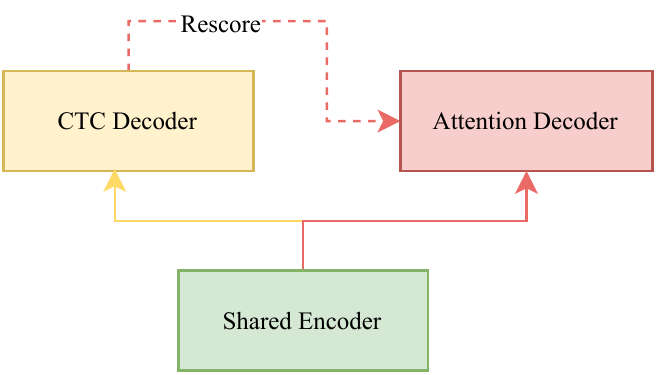}
  \caption{Two pass CTC and AED joint architecture}
  \label{fig:ctc_attention_joint}
  \vspace{-1.5em}
\end{figure}

U2, a unified two-pass joint CTC/AED model gives a good solution to the problems. As shown in Figure~\ref{fig:ctc_attention_joint}, U2 consists of three parts, \textit{Shared Encoder}, \textit{CTC Decoder} and \textit{Attention Decoder}. The \textit{Shared Encoder} consists of multiple Transformer~\cite{vaswani2017attention} or Conformer~\cite{gulati2020conformer} layers which only takes limited right contexts into account to keep a balanced latency. The \textit{CTC Decoder} consists of a linear layer, which transforms the \textit{Shared Encoder} output to the CTC activation while the \textit{Attention Decoder} consists of multiple Transformer decoder layers. During the decoding process, the \textit{CTC Decoder} runs in a streaming mode in the first pass and the \textit{Attention Decoder} is used in the second pass to give a more accurate result.

\subsubsection{Training}

The CTC loss and AED loss are combined in the training of U2:
\begin{equation}
  \vspace{-4pt}
  \label{eq:combined_loss}
    \mathbf{L}_{combined}\left(\mathbf{x}, \mathbf{y}\right)=\lambda \mathbf{L}_{\text{CTC}}\left(\mathbf{x}, \mathbf{y}\right)+(1-\lambda) \mathbf{L}_{\text{AED }}\left(\mathbf{x}, \mathbf{y}\right)
\end{equation}
where $\mathbf{x}$ is the acoustic feature, $\mathbf{y}$ is the corresponding label, $\mathbf{L}_{\text{CTC}}\left(\mathbf{x}, \mathbf{y}\right)$ and
$\mathbf{L}_{\text{AED}}\left(\mathbf{x}, \mathbf{y}\right)$ are the CTC and AED loss respectively while
$\lambda$ is a hyperparameter which balances the importance of CTC and AED loss.

As described before, our U2 can work in streaming mode when the \textit{Shared Encoder} doesn't need information of the full utterance. 
We adopt a dynamic chunk training technique to unify the non-streaming and streaming model. Firstly, the input is split into several chunks by a fixed chunk size $C$ with inputs $[t+1, t+2, ..., t+C]$ and every chunk attends on itself and all the previous chunks, so the whole latency for the \textit{CTC Decoder} in the first pass only depends on the chunk size. When the chunk size is limited, it works in a streaming way; otherwise it works in a non-streaming way.
Secondly, the chunk size is varied dynamically from 1 to the max length of the current training utterance in the training, so the trained model learns to predict with arbitrary chunk size. Empirically, a larger chunk size gives better results with higher latency, so we can easily balance the accuracy and latency by tuning the chunk size during the runtime.

\subsubsection{Decoding} \label{section:decoding_in_u2}

To compare and evaluate different parts of the joint CTC/AED model during the Python based decoding process in the research stage, WeNet supports four decoding modes as follows:
\begin{itemize}
    \item \textbf{attention}: apply standard autoregressive beam search on the AED part of the model.
    \item \textbf{ctc\_greedy\_search}: apply CTC greedy search on the CTC part of the model, CTC greedy search is super faster than other modes.
    \item \textbf{ctc\_prefix\_beam\_search}: apply CTC prefix beam search on the CTC part of the model, which can give the n-best candidates.
    \item \textbf{attention\_rescoring}: first apply CTC prefix beam search on the CTC part of the model to generate n-best candidates, and then rescore the n-best candidates on the AED decoder part with corresponding encoder output.
\end{itemize}

During the development runtime stage, WeNet supports the attention\_rescoring decodeing mode only since it's our ultimate solution for production.

\subsection{System Design}

\begin{figure}[!h]
  \vspace{-5pt}
  \centering
  \includegraphics[width=\linewidth]{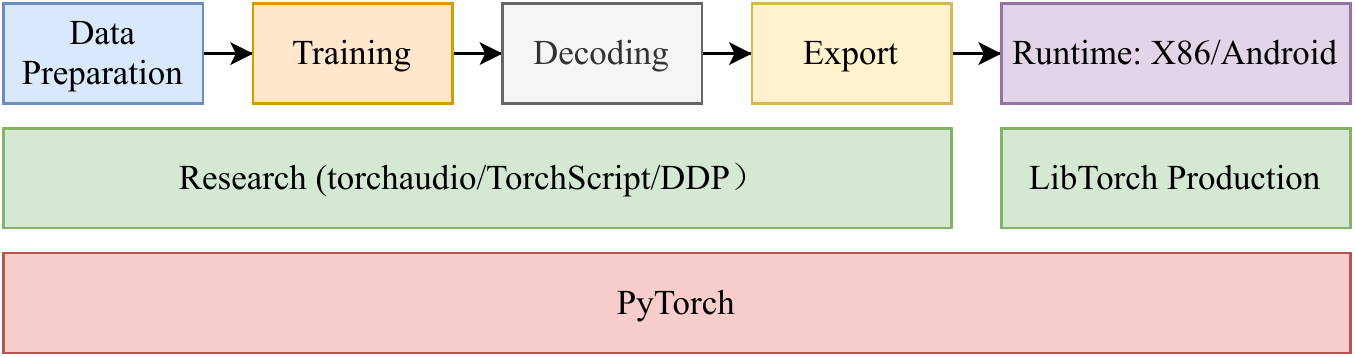}
  \caption{The design stack of WeNet}
  \label{fig:wenet_design}
  \vspace{-15pt}
\end{figure}

The overall design stack of WeNet is shown in Figure~\ref{fig:wenet_design}. Note that the bottom stack is fully based on PyTorch and it's ecosystem. The middle stack consists of two parts. When we develop a research model, TorchScript is used for developing models, Torchaudio is used for on-the-fly feature extraction, Distributed Data Parallel (DDP) is used for distributed training, torch Just In Time (JIT) is used for model exportation, PyTorch quantization is used to quantize model, and LibTorch is used for production runtime. LibTorch Production is for hosting production model, which is designed to support various hardwares  and platforms like CPU, GPU (CUDA) Linux, Android, and iOS.
The top stack shows typical research to production pipeline in WeNet. The following subsections will go through the detailed design of these modules.

\subsubsection{Data Preparation}
There is no need for any offline feature extraction in the data preparation stage since we use on-the-fly feature extraction in training. WeNet just needs a Kaldi format transcript, a wave list file and a model unit dictionary to create input file.

\subsubsection{Training}

The training stage in WeNet has the following key features. 

\textbf{On-the-fly feature extraction}: this is based on Torchaudio which can generate the same Mel-filter banks feature as Kaldi does. Since the feature is extracted on-the-fly from the raw PCM data, we can perform data augmentation on raw PCM at both time and frequency levels, and finally feature level at the same time, which enrich the diversity of data. 

\textbf{Joint CTC/AED training}: joint training speeds up the convergence of the training, improves the stability of the training as well as gives better recognition results. 

\textbf{Distributed training}: WeNet supports multiple GPUs training with DistributedDataParallel in PyTorch in order to make full use of multi-worker multi-gpu resources to achieve a higher linear speedup.

\subsubsection{Decoding}

A set of Python tools are provided to recognize wave files and compute accuracy in different decode modes. These tools help users to validate and debug the model before deploying it in production. All the decoding algorithms in Section ~\ref{section:decoding_in_u2} are supported. 

\subsubsection{Export}

As a WeNet model is implemented in TorchScript, it can be exported by torch JIT to the production directly and safely. Then the exported model can be hosted by using the LibTorch library in runtime  while both float-32 model and quantized int-8 model are supported. Using a quantized model can double the inference speed or even more when hosted on embedd devices such as ARM based Android and iOS platform.

\subsubsection{Runtime}
Currently, we support hosting WeNet production model on two mainstream platforms, namely x86 as server runtime and Android as on-device runtime. A C++ API Library and runnable demos for both platforms are provided while the users can also implement their customized system by using the C++ library. We carefully evaluated the three key metrics of an ASR system, which are accuracy, real-time factor (RTF), and latency. The results reported in Section~\ref{section:runtime_benchmark} will show that WeNet is suitable for many ASR applications including service API and on-device voice assistants.
\section{Experiments}

We carry out our experiments on the open-source Chinese Mandarin speech corpus AISHELL-1~\cite{bu2017aishell}, which contains a 150-hour training set, a 10-hour development set and a 5-hour test set. The test set contains 7,176 utterances in total. As for acoustic features, the 80-dimensional log Mel-filter banks (FBANK) are computed on-the-fly by Torchaudio with a 25ms window and a 10ms shift. Besides, SpecAugment~\cite{park2019specaugment} is applied 2 frequency masks with maximum frequency mask ($F = 10$), and 2 time masks with maximum time mask ($T=50$) to alleviate over-fitting. Two convolution sub-sampling layers with kernel size 3*3 and stride 2 are used in the front of the encoder. For model parameters, we use 12 transformer layers for the encoder and 6 transformer layers for the decoder. Adam optimizer is used with a learning rate schedule with 25,000 warm-up steps. Moreover, we obtain our final model by averaging the top-K best models which have a lower loss on the development set during training.

\subsection{Unified model evaluation}

We first evaluate a non-streaming model (M1) as our baseline which is trained and inferenced by full attention and another unified model (M2) with a dynamic chunk strategy. M2 is inferenced with different chunk sizes (full/16/8/4) at decoding, where full means full attention non-streaming case and 16/8/4 is for the streaming case.

\begin{table}[ht]
\footnotesize
\vspace{-5pt}
\caption{Unified model evaluation on AISHELL-1 test set with CER(\%).}
\label{tab:unified_model}
\centering
\scalebox{0.95}[1.0]
{
\begin{tabular}{llllll}
\toprule
\multirow{2}{*}{\textbf{decoding method}} & \multirow{2}{*}{\textbf{M1}} & \multicolumn{4}{c}{\textbf{M2}} \\ 
\cline{3-6}
                           &      & full & 16   & 8    & 4           \\ \midrule
attention                  & 5.69 & 6.04 & 6.35 & 6.45 & 6.70        \\
ctc\_greedy\_search        & 5.92 & 6.28 & 6.99 & 7.39 & 7.89       \\
ctc\_prefix\_beam\_search  & 5.91 & 6.28 & 6.98 & 7.40 & 7.89       \\
attention\_rescoring       & 5.30 & 5.52 & 6.05 & 6.28 & 6.62        \\ \bottomrule           
\end{tabular}%
}
\vspace{-6pt}
\end{table}

As shown in Table~\ref{tab:unified_model}, the unified model not only shows comparable results to the non-streaming model on the full attention case but also gives promising results on the streaming case with limited chunk size 16/8/4, which verifies the effectiveness of the dynamic chunk training strategy.

Comparing the four different decoding modes, the attention\_rescoring mode can always improve on the CTC results in both the non-streaming mode and the unified mode. The ctc\_greedy\_search and ctc\_prefix\_beam\_search have almost the same performance, and they degrade significantly as the chunk size decreases. The attention\_rescoring mode alleviates the performance degradation of the ctc\_prefix\_beam\_search results while the attention mode degrades the performance slightly.

The attention\_rescoring mode is faster and has a better RTF than the attention mode since the attention mode is an autoregressive procedure while the attention\_rescoring mode is not.
Overall, the attention\_rescoring mode not only shows promising results but also has a lower RTF. As a result, the dynamic chunk based unified model with attention\_rescoring decoding is our choice for production. So only the attention\_rescoring mode is supported at runtime.

\subsection{Runtime benchmark}
\label{section:runtime_benchmark}
This section will show the quantization, RTF, and latency benchmarks on the unified model M2 described above. We finished benchmarks on a server x86 platform and an on-device ARM Android platform respectively.

For the cloud x86 platform, the CPU is 4 cores Intel(R) Xeon(R) CPU E5-2620 v4 @ 2.10GHz with 16G memory in total. Only one thread is used for CPU threading and TorchScript inference \footnote{https://pytorch.org/docs/stable/notes/cpu\_threading\_torchscript\_inf\\erence.html} for each utterance since the cloud service requires parallel processing, and a single thread avoids performance degradation in parallel processing. For the on-device Android, the CPU is 4 cores Qualcomm Snapdragon 865 with 8G memory and a single thread is used for the on-device inference.

\subsubsection{Quantization}
Here we just compare CER before and after quantization. As shown in Table~\ref{tab:cer_on_qunantization}, CER is comparable before and after quantization. CER of the float model is slightly different from what we listed in Table~\ref{tab:unified_model} because the result in Table~\ref{tab:unified_model} is evaluated by Python research tools while that in Table~\ref{tab:cer_on_qunantization} here is evaluated by runtime tools.

\begin{table}[h]
\vspace{-5pt}
\footnotesize
\caption{CER before and after quantization on AISHELL-1 test set. All results are evaluated using runtime tools.}
\vspace{-4pt}
\label{tab:cer_on_qunantization}
\centering
\begin{tabular}{@{}lllll@{}}
\toprule
quantization/decoding\_chunk & full  & 16    & 8     & 4  \\ \midrule
NO (float32)                    & 5.58 & 6.03 & 6.27 & 6.60  \\
YES (int8)                    & 5.59 & 6.06 & 6.28 & 6.64  \\ \bottomrule
\end{tabular}%
\vspace{-20pt}
\end{table}

\subsubsection{RTF}

As shown in Table~\ref{tab:rtf_benchmark}, RTF increases as the chunk size decreases since smaller chunk requires more iterations for forward computation. Further, quantization can bring about 2 times speedup on on-device (Android) and a slight improvement on server (x86).

\begin{table}[ht]
\footnotesize
\vspace{-8pt}
\caption{RTF benchmark}
\vspace{-5pt}
\label{tab:rtf_benchmark}
\centering
\begin{tabular}{@{}lllll@{}}
\toprule
model/decoding\_chunk    & full  & 16    & 8     & 4     \\ \midrule
server (x86) float32        & 0.079 & 0.095 & 0.128 & 0.186 \\
server (x86) int8         & 0.072 & 0.081 & 0.098 & 0.134 \\
on-device (Android) float32 & 0.164 & 0.251 & 0.350 & 0.505 \\
on-device (Android) int8  & 0.082 & 0.114 & 0.130 & 0.201 \\ \bottomrule
\end{tabular}%
\vspace{-14pt}
\end{table}

\subsubsection{Latency}

For latency benchmark, we create a WebSocket server/client to simulate a real streaming application because this benchmark is only carried out on the server x86 platform. The average latency we evaluated is described here.
\textbf{Model latency (L1)}: the waiting time introduced by the model structure. For our chunk based decoding, the average waiting time is half of the chunk theoretically. And the total model latency of our model is $(chunk/2*4+6)*10$ (ms), where 4 is the subsampling rate, 6 is the lookahead introduced by the first two CNN layers in the encoder and 10 is the frame shift.
\textbf{Rescoring cost (L2)}: the time cost on the second pass attention rescoring.
\textbf{Final latency (L3)}: the user (client) perceived latency, which is the time difference between the user stopping speaking and getting the recognition result. When our ASR server receives the speech ending signal, it first forwards the left speech for CTC searching, and then does the second pass attention rescoring, so rescoring cost is part of the final latency. The network latency also should be taken into account for a real production but it is negligible, since we test the server/client on the same machine.

\begin{table}[ht]
\footnotesize
\vspace{-5pt}
\caption{Latency benchmark on the server x86 platform with simulated WebSocket server/client.}
\vspace{-1pt}
\label{tab:latency_benchmark}
\centering
\begin{tabular}{@{}llll@{}}
\toprule
decoding\_chunk & \textbf{L1} (ms) & \textbf{L2} (ms) & \textbf{L3} (ms) \\ \midrule
16              & 380    & 115  & 142    \\
8               & 220    & 115  & 135    \\
4               & 140    & 114  & 130    \\ \bottomrule
\end{tabular}%
\vspace{-10pt}
\end{table}

As shown in Table~\ref{tab:latency_benchmark}, the rescoring costs are almost the same for different chunk sizes and this is reasonable since rescoring computation is invariant to chunk size.
Besides, the final latency is dominated by the rescoring cost which means we can further reduce the final latency by reducing the rescoring cost. At last, the final latency increases slightly as the decoding chunk varies from 4 to 8 and from 8 to 16.

\subsection{15,000-hour Task}

\begin{table}[t]
\footnotesize
    \vspace{-10pt}
    \centering
    \setlength{\abovecaptionskip}{0.1cm}
    \caption{15,000-hour Mandarin speech data trained U2 and static full attention (Conformer), tested on three test sets in CER (\%). Utt Dur (s) denotes average utterance duration in seconds.}
    \label{tab:15000 hour}
    \begin{tabular}{c c c c c c}
    \toprule
    \multirow{2}{*}{test set}  &   \multirow{2}{*}{Utt Dur (s)} &
    \multirow{2}{*}{Conformer} & \multicolumn{2}{c}{U2} \\ \cline{4-5} 
       & &  & full   & 16          \\ \midrule
    \multirow{1}{*}{AISHELL-1} & 5.01
                                                                    & 3.96            & 3.70   & 4.41    \\ 
    \multirow{1}{*}{TV}  & 2.99  & 10.96           & 11.61   & 13.03    \\ 
    \multirow{1}{*}{Conversation}  & 2.69  & 12.84            & 13.86   & 15.07    \\ \bottomrule
    \end{tabular}%
    \vspace{-15pt}
\end{table}

We further train the proposed U2 model using a 15,000-hour Mandarin dataset collected from various domains which include talk show, TV play, podcast and radio broadcast to show our model's ability on an industry-scale dataset. The model is evaluated on three test sets.

We use Conformer~\cite{gulati2020conformer} as our shared encoder while the decoder is a transformer as same as the previous experiments. Conformer adds convolution module on the basis of transformer so it can capture both local and global context and get better results on different ASR tasks. Specially, the causal convolution is used  for chunk training of Conformer and we add extra 3 dimensional pitch features concatenated with 80-dimensional FBANK. We keep the main structure of the encoder in previous experiments and only change the transformer layers to 12 conformer layers with multi-head attention (4 heads). Each conformer layer uses 384 attention dimension and 2048 feed forward dimension. 
Besides, accumulating grad is also used to stabilize training and we update parameters every 4 steps. Moreover, we obtain our final model by averaging the top-10 best models which have a lower loss on the evaluation set during training. We train a full context conformer CTC model and a U2 model using dynamic chunk training. Three test sets are used to evaluate these models, including AISHELL-1, TV domain and conversation domain. The U2 model works on attention\_rescoring decoding mode while the conformer model works on the attention decoding mode.
As shown in Table~\ref{tab:15000 hour}, we can see that U2 achieves comparable results with the Conformer baseline in general and even better results on AISHELL-1 test set when using full attention during inference. When the chunk size is 16, the CER does not get obviously worse. 

To analyse why U2 model perform better on the AISHELL-1 task, we collect the average utterance duration in each test set shown in Table~\ref{tab:15000 hour}. Since the average utterance duration in AISHELL-1 is much longer than that in the other two test sets, AISHELL-1 task may need stronger global information modeling ability. The U2 model can use the attention decoder to rescore the CTC hypotheses, which makes it more favourable to the AISHELL-1 task.

\section{Conclusions}
We present a new open source production oriented E2E speech recognition toolkit named WeNet, providing a unified solution for streaming and non-streaming application. This paper introduces the model structure behind the toolkit, system design and benchmarks. The whole toolkit is well designed, lightweight and shows great performance on an open dataset and an internal large dataset. Wenet has already supported custom language model in runtime, by adopting n-gram ad WFST. In addition, Wenet has also supported gRPC-based speech recognition microservice framework applications. More featured functions will be updated in the future. Please stay tuned and visit our website https://github.com/wenet-e2e/wenet for more updates.

\newpage

\bibliographystyle{IEEEtran}
\bibliography{main}

\end{document}